\documentclass[pre,aps,floatfix,showpacs]{revtex4-1}

\usepackage{bm}
\usepackage{amsmath}
\usepackage{amssymb}

\begin{document}
\title{Scaling the universe}

\author{Norman E. Frankel}
\email{nef@unimelb.edu.au}

\affiliation{School of Physics, University of Melbourne,
Victoria 3010, Australia}

\pacs{04.40.-b, 05.30.-d, 05.45.Df, 98.65.-r}
\date{\today}

%% Abstract
\begin{abstract}
A model is presented for the origin of the large scale structure of the universe and
their Mass-Radius scaling law; a fractal power law, $M \propto R^D$,
with dimension $D=2$, most significantly.
The physics is conventional, orthodox, but it is used to
fashion a highly unorthodox model of the origin of the galaxies, their groups, clusters,
super-clusters, and great walls. The scaling law fits the observational results and the
model offers new suggestions and predictions. These include a largest, a supreme,
cosmic structure, and possible implications for the recently observed pressing cosmological
anomalies.
\end{abstract}

\maketitle
%%%%%%%%%%%%%%%%%%%%%%%%%%%%%%%%%%%%%%%%%%%%%%%%%%%%%%%%%

\section{Introduction}
\label{sect1}

One of the most compelling problems in cosmology is the origin and evolution of
the large scale structure of the universe. These heavenly bodies, the galaxies and
their groups, clusters, super-clusters and great walls, are as enigmatic as they are
alluring.

The current scenario decrees that the tiny density fluctuations, reflected in the
tiny temperature variations in the cosmic microwave background (CMB), act as
the seeds which grow, under gravitational attraction as the universe expands, into
 these bodies. See the website ~\cite{1} 
%(http//map.gsfc.nasa.gov/universe/bb\_cosmo\_struct.html)
for a pr$\acute {\rm e}$cis of the current picture. We'll say
no more on this at this point, save to note that there is no detailed 
analytical theory,
akin to that for stellar structure, presently accompanying this picture.
Later we will 
discuss the physics in this paper along with the problems of 
the usual interpretation with regard to the pressing anomalies 
in the latest experimental results.

One of the key signatures of such a model theory should be the Mass-Radius
scaling law for the large scale structure. 
We find such a scaling law which very accurately fits the Mass-Radius of
all the assemblies in the large scale structure.
The Mass-Radius scaling law is a fractal scaling law of the
form  $M \propto R^D$ with the exponent $D = 2$, which is most
significant.

In the atomic theory of matter, taking
hydrogen as the prototype atom, we know that the energy scales as the Rydberg
$= \frac{1}{2} m_e c^2 (\alpha_e)^2$ and the size scales as the Bohr radius, 
$a_o = \lambda_e/(\alpha_e)$. Here $m_e$ and
$\lambda_e$ are the electron's mass and Compton wavelength, and 
$\alpha_e = e^2/(\hbar c)$ is the fine structure constant.

In astrophysics, thanks to the seminal works of Eddington~\cite{2}, 
Fowler~\cite{3} and Chandrasekhar~\cite{4},
we know that the structure of stellar matter (stars) have masses and
radii that scale as shown in Table~\ref{tab1}, where $\alpha_G = GH^2/(\hbar c)$ 
is the gravitational fine structure constant. Here, in honour of Eddington, we have 
chosen to represent, as he did, the mass of the nucleon as H.
%%%%%%%%%%%%%%%%%%%%%%%%%%%%%%%%%%%%%%%%%%%%%%%%%%%%%%%
\begin{table}[h]
\caption{Stellar Scaling.}
\begin{ruledtabular}
\begin{tabular}{ccc} 
Mass & Radius & Structure\\ 
\hline
 & \hspace*{0.4cm}
$ \left(\frac{1}{\alpha_G} \right)^\frac{1}{2} a_o$ \hspace*{0.4cm} 
&\hspace*{0.4cm} 
 Sun \hspace*{0.4cm}\\
$\left( \frac{1}{\alpha_G}\right)^\frac{3}{2} H $&
$\left(\frac{1}{\alpha_G} \right)^\frac{1}{2} \lambda_e$ & White Dwarf\\
 &$\left(\frac{1}{\alpha_G} \right)^\frac{1}{2} \lambda_H$ & Neutron Star\\
\end{tabular}
\label{tab1}
\end{ruledtabular}
\end{table}
%%%%%%%%%%%%%%%%%%%%%%%%%%%%%%%%%%%%%%%%%%%%%%%%%%%%%%%%%%%%%%%

The masses of the white dwarf and neutron stars in Table~\ref{tab1} 
are their critical masses, as obtained from the ultra-degenerate, ultra-relativistic, 
stellar structure theory. Nowadays, we can see this simply for all three classes of 
stars, where their internal pressure is  $P_i \propto  \hbar c(\rho/H)^\frac{4}{3}$,
with $\rho$ their mass density. When balanced against
the gravitational pressure, $P_G \propto GM\rho/R$, this readily gives their 
[critical] mass, $M \propto (1/\alpha_G)^\frac{3}{2} H$
as is exhibited in Table~\ref{tab1}.

The theory of white dwarfs owes its inception to Fowler~\cite{3} who established that theory 
in his classic paper, following directly upon the independent discovery
by Fermi and Dirac, of the quantum statistical mechanics of ideal fermions (as they
are now called). Following on from Fowler, nowadays we can readily find for 
ultra-degenerate, non-relativistic, white dwarf and neutron stars, this utterly remarkable,
beautiful, compact Mass-Radius scaling law,
%%%%%%%%%%%%%%%%%%%%%%%%%%%%%%%%%%%%%%%%%%%%%%%%%%%%%%%%%%%%%%%%
\begin{equation}
\left({\overline M}\right)^\frac{1}{3} \ {\overline R}\ 
= C = 4.51227..... , 
\end{equation}
%%%%%%%%%%%%%%%%%%%%%%%%%%%%%%%%%%%%%%%%%%%%%%%%%%%%%%%%%
where ${\overline M} = M/M_o$ with $M_o = (1/\alpha_G)^\frac{3}{2} H$
and ${\overline R} = R/R_o$ with $R_o =(1/\alpha_G)^\frac{1}{2}\lambda_o$,
where $\lambda_o$ is the Compton wavelength of the electron/neutron for the white
dwarf/neutron star. This law holds for non-relativistic white dwarfs,
requiring the Fermi momentum, $p_F < m_ec$ , meaning that
%%%%%%%%%%%%%%%%%%%%%%%%%%%%%%%%%%%%%%%%%%%%%%%%%%%%%%%%%
\begin{equation}
\rho < \frac{1}{3\pi^2} \frac{H}{\lambda_e^3} .
\end{equation}
%%%%%%%%%%%%%%%%%%%%%%%%%%%%%%%%%%%%%%%%%%%%%%%%%%%%%%%%%%%%%%%%
This Fowler Scaling Law has been an abiding inspiration in our search for a model
of, and concomitant scaling law for, the large scale structure of the universe.

There have been two prime motivational insights that have guided our search.
The first is the utterly remarkable numerical coincidence, first recognised by de
Sitter and Weyl and Eddington (see the nice review in Ref.~\cite{5}), that
%%%%%%%%%%%%%%%%%%%%%%%%%%%%%%%%%%%%%%%%%%%%%%%%%%%%
\begin{equation}
\frac{\rm a\ cosmic\ mass}{\rm a\ particle\ mass}  \propto
\frac{1}{\alpha_G^2}
\end{equation}
%%%%%%%%%%%%%%%%%%%%%%%%%%%%%%%%%%%%%%%%%%%%%%%%%%%%%%%%%%%%%%%
and
%%%%%%%%%%%%%%%%%%%%%%%%%%%%%%%%%%%%%%%%%%%%%%%%%%%%%%%%%%
\begin{equation}
\frac{\rm a\ cosmic\ size\ -\ radius}
{\rm a\ particle\ size}  \propto \frac{1}{\alpha_G} .
\end{equation}
%%%%%%%%%%%%%%%%%%%%%%%%%%%%%%%%%%%%%%%%%%%%%%%%%%%%%%%%%%
This correspondence was prosecuted relentlessly by Eddington throughout his 
writings~\cite{6,7,8}  and has been a persistent inspiration to us throughout our long 
search.

The second is the pioneering work of Peebles~\cite{9} ,
followed up by many since, who studied the mass distributions of the large scale structure, 
by analyzing [the now extensive] data, using the 2-point, galaxy-galaxy, correlation 
function, $\zeta(r) \propto r^{-p}$.
There is now very strong evidence~\cite{10,11,Ba98,La98} 
that the exponent $p = 1$, thus indicating a mass proportional to radius squared relationship.

These two, seemingly disparate, insights in fact are intimately connected as will
be seen in the following sections in which we will develop a model of, and scaling law
for, the large scale structure of the universe. Predictions and possible implications
for the pressing cosmological anomalies will also be given.

\section{The Dawn of Structure --- Part I}
\label{sect2}

We turn our attention to the era just prior to recombination (decoupling). This
period corresponds to temperature, $T_r$, of the universe, which given by the Saha
equation is $T_r = 3000$ K. The current temperature of the universe is $T_c = 2.725$ K,
so we know this era corresponds to a red-shift $z_r = T_r/T_c = 1100$. [Throughout we
use z for z +1]

We will also need the era when the matter contribution to the universe's mass
density $\rho_m$ equals $\rho_\gamma$,the radiation contribution to that mass density.
Cosmological theory and experiment~\cite{12}
both give this as $z \simeq 3300$. So at the era just before
recombination, we have $\rho_m \simeq (3300/1100) \rho_\gamma = 3 \rho_\gamma$. 
The total mass density of the
universe then is $\rho = \rho_m + \rho_\gamma = 4 \rho_\gamma$. 
Everything now can be computed from the
physical quantities of the radiation (photons and neutrinos).

Proceeding, the Jeans stability condition [see also the appendix] is
%%%%%%%%%%%%%%%%%%%%%%%%%%%%%%%%%%%%%%%%%%%%%%%%%%%%%%
\begin{equation}
w^2 = -4\pi G \rho + k^2 v^2 , 
\end{equation}
%%%%%%%%%%%%%%%%%%%%%%%%%%%%%%%%%%%%%%%%%%%%%%%%%%%%%%%%%%%%%%
where $\rho$ is the total mass density, $k$ is the wave number, and $v$ is the speed 
of sound $v^2  = (\partial P)/(\partial \rho)$, where $P$ is the total pressure of the 
universe, $P = P_m + P_\gamma$.

The Jeans mass, $M_J$, and the Jeans radius, $R_J$, at this era are readily found.
The radius is
%%%%%%%%%%%%%%%%%%%%%%%%%%%%%%%%%%%%%%%%%%%%%%%%%%%%%%%%%%%%%%%%%%%
\begin{equation}
R_J^2 = \frac{5}{16\pi} \left(\frac{1}{\alpha_G} \right)
\left(\frac{T_H}{T_r} \right)^4 \lambda_H^2 \ F^{-1}
\end{equation}
%%%%%%%%%%%%%%%%%%%%%%%%%%%%%%%%%%%%%%%%%%%%%%%%%%%
 where
%%%%%%%%%%%%%%%%%%%%%%%%%%%%%%%%%%%%%%%%%%%%%%%%%%%%%
\[
T_H = \frac{Hc^2}{k} = 1.0888..... \times 10^{13} {\rm K}.
\]
%%%%%%%%%%%%%%%%%%%%%%%%%%%%%%%%%%%%%%%%%%%%%%%%%%%%
and
%%%%%%%%%%%%%%%%%%%%%%%%%%%%%%%%%%%%%%%%%%%%%%%%%%%%%%%
\begin{displaymath}
F = \left[1 + 3.046 \times \frac{7}{8} \times 
\left( \frac{4}{11} \right)^{\frac{4}{3}} \right] = 1.6918 ......
\end{displaymath}
%%%%%%%%%%%%%%%%%%%%%%%%%%%%%%%%%%%%%%%%%%%%%%%%%%%%%%%
in which `$1$' comes from the photons while the second term accounts for the
neutrinos (see Eq.~(1) in \cite{37}).
Here $k$ is the Boltzmann constant and we use
%%%%%%%%%%%%%%%%%%%%%%%%%%%%%%%%%%%%%%%%%%%%%%%%%%%%%%%%%%%
\begin{equation}
\rho = \frac{4\pi^2}{15} \frac{\left(kT_r\right)^4}{\hbar^3 c^5}\ F
= \frac{4\pi^2}{15} \left(\frac{T_r}{T_H}\right)^4 
\frac{H}{\lambda_H^3}\ F.
\end{equation}
%%%%%%%%%%%%%%%%%%%%%%%%%%%%%%%%%%%%%%%%%%%%%%%%%%%%%%%%%%%
Readily it is shown that
%%%%%%%%%%%%%%%%%%%%%%%%%%%%%%%%%%%%%%%%%%%%%%%%%%
\begin{equation}
\frac{P_\gamma}{P_m} \simeq \frac{1}{9} \left(\frac{T_H}{T_r} \right) ,
\end{equation}
%%%%%%%%%%%%%%%%%%%%%%%%%%%%%%%%%%%%%%%%%%%%%%%%%%
so that the total pressure is $P = \frac{1}{3} \rho_\gamma c^2$ and 
$v^2 = c^2/12$. The Jeans mass is
%%%%%%%%%%%%%%%%%%%%%%%%%%%%%%%%%%%%%%%%%%%%%%%%%%%%%%%%%
\begin{equation}
M_J = \frac{4\pi}{3} \rho R_J^3
= \frac{\pi}{36} \sqrt{5\pi} \left(\frac{1}{\alpha_G}\right)^\frac{3}{2}
\left(\frac{T_H}{T_r}\right)^2 H\ F^{-\frac{1}{2}} .
\end{equation}
%%%%%%%%%%%%%%%%%%%%%%%%%%%%%%%%%%%%%%%%%%%%%%%%%%%%%%%%
Now, for the `pi$\grave {\rm e}$ce de r$\acute{\rm e}$sistance', 
we find this utterly remarkable numerical coincidence,
%%%%%%%%%%%%%%%%%%%%%%%%%%%%%%%%%%%%%%%%%%%%%%%%%%%%%%%%%
\begin{equation}
\frac{T_H}{T_r}  
= 1.0061..\left(\frac{1}{\alpha_G}\right)^\frac{1}{4}
= 1.0061..\left(1.6933 \times 10^{38}\right)^\frac{1}{4} .
\end{equation}
%%%%%%%%%%%%%%%%%%%%%%%%%%%%%%%%%%%%%%%%%%%%%%%%%%%%%%%%%%
Such numerical coincidences have solid history, having been advanced first 
by the insightful works~\cite{13,14,15}. 
See also the fine review with references~\cite{16}
and the nice book~\cite{17}.  Of course, Eddington was already
on the trail of such, several decades earlier.

Bolstered by this most fortunate of coincidences, we use
%%%%%%%%%%%%%%%%%%%%%%%%%%%%%%%%%%%%%%%%%%%%%%%%%%%%%%%%%%%%%%%%%%
\begin{equation}
\left(\frac{T_H}{T_r} \right)^4 = \left(\frac{1}{\alpha_G}\right) .
\end{equation}
%%%%%%%%%%%%%%%%%%%%%%%%%%%%%%%%%%%%%%%%%%%%%%%%%%%%%%%%%%%%
This correspondence is utterly essential in establishing a fundamental scaling law
for the large scale structure of the universe. We now have at this era
%%%%%%%%%%%%%%%%%%%%%%%%%%%%%%%%%%%%%%%%%%%%%%%%%%%%%%%%%%%%%%%
\begin{align}
R_J =& \sqrt{\frac{5}{16\pi}} \left(\frac{1}{\alpha_G} \right) \lambda_H
\ F^{-\frac{1}{2}} \\
\rho\ \ =& \frac{4\pi^2}{15} \alpha_G \frac{H}{\lambda_H^3}\ F\\
M_J =& \frac{\pi}{36} \sqrt{5\pi} \left( \frac{1}{\alpha_G}\right)^2 H\
F^{-\frac{1}{2}}\  
= \sigma_J R_J^2\\
\sigma_J =& \frac{4\pi^2}{45} \sqrt{5\pi} \frac{H}{\lambda_H^2}\ F^{\frac{1}{2}} .
\label{Eq15}
\end{align}
%%%%%%%%%%%%%%%%%%%%%%%%%%%%%%%%%%%%%%%%%%%%%%%%%%%%%
Most satisfying, we now have the scalings,
%%%%%%%%%%%%%%%%%%%%%%%%%%%%%%%%%%%%%%%%%%%%%%
\begin{align}
\frac{M_J}{H} &\propto \left(\frac{1}{\alpha_G}\right)^2 \\
\frac{R_J}{\lambda_H} &\propto \left(\frac{1}{\alpha_G}\right) \\
M_J &\propto R_J^2 .
\end{align}
%%%%%%%%%%%%%%%%%%%%%%%%%%%%%%%%%%%%%%%%%%%%%%%%%%%

In closing this section, and leading to the next, we remark that the very important 
work of Chandrasekhar showed that such a giant ultra-radiation dominated structure as
we have here, must be unstable and collapse. That was found to be the
case independently by Feynman, and it will be discussed in Sec.~\ref{sect4}. This collapse is
required, of course, and is essential for the continuing development of our model and
quest for a fundamental scaling law for the large scale structure. The Jeans surface
mass density, $\sigma_J \propto H/\lambda_H^2$ gives the clue on how to proceed.

\section{The Dawn of Structure --- Part II}
\label{sect3}

The quantity $H/\lambda_H^2$ is precisely the kind of mass density to be expected for an
ultra-planar configuration (an ultra-collapsed 3-dimensional structure) of 
ultra-degenerate fermion (nucleon) matter. We envision, therefore, that the collapse of the
proto-structure described in Sec.~\ref{sect2} resulted in such a planar structure of  
ultra-degenerate neutron matter; a kind of cosmic `pancake'. The collapse to density 
$H/\lambda_H^2$ is more than sufficient to ensure that the protons and electrons in the 
proto-structure will undergo inverse $\beta$-decay forming neutrons.

To see what the mass and radius of such an ultra-planar, ultra-degenerate neutron
matter structure would be, we first look at a heuristic calculation, The statistical 
mechanics of such a system is elementary. In the ultra-relativistic limit, the
internal pressure is $P_i \propto \hbar c\left(\sigma/H\right)^\frac{3}{2}$
where the mass density $\sigma > H/\lambda_H^2$. Balancing this against the pressure 
of gravity, $P_G \propto GM\sigma/R$, we find a critical mass
$M \simeq \left(1/\alpha_G\right)^2 H$. This is remarkably satisfying as it is comparable 
to $M_J$ for the collapsed proto-system considered in Sec.~\ref{sect2}.

In the ultra-nonrelativistic region, the internal pressure is $P_i \propto
\hbar^2\sigma^2 /H$, 
and balancing this against $P_G$ results in a critical radius, in a maximum sense, of the
planar structure, $R \simeq \left(1/\alpha_G\right)\lambda_H$. This is remarkably comparable 
to the $R_J$ given in Sec.~\ref{sect2}.

Thus the collapsed proto-structure fits very comfortably in such an ultra-planar, 
ultra-degenerate, cosmological `pancake'. As the surface density of the collapsed 
proto-structure is $\sigma_J \simeq 4.52 H/\lambda_H^2$, see Eq.~(\ref{Eq15}),  
the ultra-planar 
`pancake'. is an ultra-relativistic, ultra-degenerate neutron structure since that would 
require the Fermi momentum of the neutrons to be $p_F > H c$ 
implying $\sigma >  \frac{1}{2\pi} H/\lambda_H^2$. 

We can solve exactly for the structure in a way analogous to the classic work
of Chandrasekhar for a 3-dimensional, ultra-degenerate structure. The equation of
state, $P(\sigma)$, can be calculated exactly. It is a somewhat involved algebraic entity
but we require only its derivative which turns out to be pleasingly simple, viz.
%%%%%%%%%%%%%%%%%%%%%%%%%%%%%%%%%%%%%%%%%%%%%%%%%%%%%%%
\begin{equation}
\frac{dP}{d\sigma} = \frac{1}{2} \frac{{\overline \sigma}c^2}
{\sqrt{{\overline \sigma} + 1}}
\end{equation}
%%%%%%%%%%%%%%%%%%%%%%%%%%%%%%%%%%%%%%%%%%%%%%%%%%%%%%%%%%%%%%%%%%%%%%
where ${\overline \sigma} = \sigma/\sigma_o$
and $\sigma_o = H/\{2\pi \lambda_H^2\}$ .
The scaling density, $\sigma_o$, is that for $p_F = Hc$. 
The equation for hydrostatic equilibrium (HE) can be readily constructed. We
use only the leading term for the gravitational potential for a planar configuration 
of matter as it carries the overwhelming strength of the potential. The exact
equation for HE, in scaled form for ${\overline \sigma}(y)$ is,
%%%%%%%%%%%%%%%%%%%%%%%%%%%%%%%%%%%%%%%%%%%%%%%%%%%%
\begin{equation}
\frac{1}{y} \frac{d}{dy}
\left[ y^2 \frac{1}{\sqrt{{\overline \sigma}(y) + 1}}
\frac{d{\overline \sigma}(y)}{dy}\right]
= - {\overline \sigma}(y)
\end{equation}
%%%%%%%%%%%%%%%%%%%%%%%%%%%%%%%%%%%%%%%%%%%%%%%%%%%%
It is a modified Lane-Emden equation which can be solved numerically. The radius
of the structure is $R = r_o y_o$ where
%%%%%%%%%%%%%%%%%%%%%%%%%%%%%%%%%%%%%%%%%%%%%%%
\begin{equation}
r_o = \frac{c^2}{4\pi G \sigma_o} = \frac{1}{2} 
\left(\frac{1}{\alpha_G} \right)\lambda_H ,
\end{equation}
%%%%%%%%%%%%%%%%%%%%%%%%%%%%%%%%%%%%%%%%%
with $y_o$ being the solution of ${\overline \sigma}(y_o)= 0$. The solution and 
${\overline \sigma}^\prime(y_o)$  are obtained for
varying values of ${\overline \sigma}(0)$, the central value of the scaled mass density 
${\overline \sigma}(y)$. The mass of the structure is
%%%%%%%%%%%%%%%%%%%%%%%%%%%%%%%%%%%%%%%%%%%%%%%%%%%%%%%%%%%%%%%%%%%%%%%%
\begin{equation}
M = \frac{1}{4} \left(\frac{1}{\alpha_G} \right)^2 
H\left[-y_o^2{\overline \sigma}^\prime(y_o)\right] .
\end{equation}
%%%%%%%%%%%%%%%%%%%%%%%%%%%%%%%%%%%%%%%%%%%%%%%%%%%%%%%%%%%%
While we solve these equations numerically for all values of the central density,
${\overline \sigma}(0)$,the ultra-relativistic and ultra-nonrelativistic cases are
of prime interest here; the solutions for which are easier. 
Taking, $\sigma(y) = \lambda \theta^n(y)$, 
the HE equation reduces to a Lane-Emden form,
%%%%%%%%%%%%%%%%%%%%%%%%%%%%%%%%%%%%%%%%%%%%%%%%%%%%%%%%%%%%%%%%%%%%%%%%
\begin{equation}
\frac{1}{y} \frac{d}{dy} \left[ y^2 \frac{d\theta}{dy}\right] = -\theta^n(y) ;
\end{equation}
%%%%%%%%%%%%%%%%%%%%%%%%%%%%%%%%%%%%%%%%%%%%%%%%%%%%%%%%%%%%%%%%%%
$n$ = 1 and 2 correspond to the ultra-nonrelativistic and ultra-relativistic cases
respectively. For the former, where $\lambda/\sigma_o << 1$, we find the critical radius 
to be
%%%%%%%%%%%%%%%%%%%%%%%%%%%%%%%%%%%%%%%%%%%%%%%%%%%%%%%%%%
\begin{equation}
R =1.83 \left(\frac{1}{\alpha_G} \right) \lambda_H ,     
\label{eq24}
\end{equation}
%%%%%%%%%%%%%%%%%%%%%%%%%%%%%%%%%%%%%%%%%%%%%%%%%%%%%%%%
and, for the latter, where $\lambda/\sigma_o >> 1$, we find the critical mass to be
%%%%%%%%%%%%%%%%%%%%%%%%%%%%%%%%%%%%%%%%%%%%%%%%%%%%%%%%
\begin{equation}
M =1.40 \left(\frac{1}{\alpha_G} \right)^2 H . 
\label{eq25}
\end{equation}
%%%%%%%%%%%%%%%%%%%%%%%%%%%%%%%%%%%%%%%%%%%%%%%%%%%%%
These are the exact results confirming their forms obtained from the heuristic 
analysis.

It is important to emphasise that,
in a planar configuration of ultra-degenerate matter, we find there is
both a critical mass as given in Eq.~(\ref{eq25}) and a critical radius
as given in Eq.~(\ref{eq24}). In contrast with the 3-dimensional 
case, where there is only, albeit the renown, critical mass 
as discussed in the Introduction and exhibited in Table~\ref{tab1} .   

Here again we have the crucial scaling we observed for the proto-structure,
%%%%%%%%%%%%%%%%%%%%%%%%%%%%%%%%%%%%%%%%%%%%%%%%%%%%%
\begin{equation}
\frac{M}{H} \propto \left( \frac{1}{\alpha_G}\right)^2
\hspace*{1.0cm} {\rm and}\hspace*{1.0cm}
\frac{R}{\lambda_H} \propto \left( \frac{1}{\alpha_G}\right) ,
\end{equation}
%%%%%%%%%%%%%%%%%%%%%%%%%%%%%%%%%%%%%%%%%%%%%%%%%%%%%%%
and affirm that the mammoth, ultra-radiation dominated, proto-structure on collapse 
can be easily accommodated in a mammoth, ultra-planar structure of ultra-degenerate 
neutron matter.

Of course,this structure inherently is unstable and must fragment, so that, as the
universe expands beyond this ($z_r = 1100$) era, the neutrons can then $\beta$-decay to
free protons and electrons that will go on to form atoms,that make the stars,which
ultimately assemble into the large scale structure of the universe.

How we envision the final form of our model, and are led to the ultimate goal,the
scaling for the large scale structure, is the topic of Sec.~\ref{sect5}, before which 
we discuss general relativistic considerations that are essential to it.

In closing this section, we reflect on some past works of ours that bear on the
matters covered in this section. Some 30 years ago we gave a preliminary discussion
of a planar-pancake-ultra-degenerate structure scenario of the
universe~\cite{18}, 
and showed that it would have a critical mass 
$\propto \left(\frac{1}{\alpha_G}\right)^2 H$.
It was not the picture envisioned in this section. We know
now just what it was telling us. It is important to emphasise that the planar structure
considered here is an ultra-anisotropic 3-dimensional structure where the gravitational
potential is still of the $1/r$ form. In a purely mathematical 2-dimensional space, we
know from our general relativistic studies of gravitation in such a
space~\cite{19,20,21} that there is no Newtonian limit.

\section{General Relativistic Considerations}
\label{sect4}
Eddington developed the theory of stellar structure as presented in his seminal
treatise~\cite{2}. It is the standard model of stellar structure; one of the
great achievements in modern physics. Chandrasekhar, in his set of 
lectures~\cite{22}
quite rightly described Eddington as the finest astrophysicist of his time.

The standard model, with its equation of state $P \propto \rho^\frac{4}{3}$ 
[see the start of Sec~\ref{sect1}]
gives the mass of a star as
%%%%%%%%%%%%%%%%%%%%%%%%%%%%%%%%%%%%%%%%%%%%%%%%%%%%%%%%%%%%%%
\begin{equation}
M = 18.1 \sqrt{\left[ \frac{(1-\beta)}{\beta^4} \right]} \ M_\odot ,
\label{starM}
\end{equation}
%%%%%%%%%%%%%%%%%%%%%%%%%%%%%%%%%%%%%%%%%%%%%%%%%%%%%%%%%%%%%
where $\beta  = P_m/P$, Here $P_m$ is, as in Sec.~\ref{sect2}, the contribution of 
matter (particles) to the total $P = P_m + P_\gamma$, where, also as in 
Sec.~\ref{sect2}, 
$P_\gamma$ is the contribution of radiation (photons).

It was Chandrasekhar, who from the outset in his seminal work~\cite{4} 
on stellar structure, recognised and emphasized that it was the combination of 
fundamental
constants of nature in the result, such as $\left(\frac{1}{\alpha_G}\right)^\frac{3}{2} H$
that characterised the mass of stars.
Thus Eq.~(\ref{starM}) can be  found to be,
%%%%%%%%%%%%%%%%%%%%%%%%%%%%%%%%%%%%%%%%%%%%%%%%%%%%%%%%%%%%%%%%
\begin{equation}
M = 9.77 \ \sqrt{\left[ \frac{(1-\beta)}{\beta^4} \right]}
\left( \frac{1}{\alpha_G} \right)^\frac{3}{2}\ H .
\end{equation}
%%%%%%%%%%%%%%%%%%%%%%%%%%%%%%%%%%%%%%%%%%%%%%%%%%%%%%%%%%%%%%%%
For our mammoth, ultra-radiation dominated, structure of Sec.~\ref{sect2}, 
where $\beta = 9(\alpha_G)^\frac{1}{4}$,
we have $M = 0.12 \left(\frac{1}{\alpha_G}\right)^2 H$, recovering the correct scaling 
as required for $M_J$ given in Sec.~\ref{sect2}.

To understand why this structure must be unstable and collapse, we turn back
in time, to the work of Hoyle and Fowler~\cite{23}
who were the first to consider radiation
dominated stars. They proposed stars of mass of the order of $10^6 M_\odot$, as a model
for the then newly discovered quasars. Though they were not able to provide for
exactly how such stars would appear, their study was nonetheless very interesting,
providing stimulating results and suggestions.

The main reason for their nonexistence came at the same time from 
Chandrasekhar~\cite{24,25,26} 
in his elegant calculations which showed that such massive radiation-
dominated stars are unstable due to general relativistic effects, and thus the stars
would collapse. Feynman, simultaneously and independently, suggested the same for
the same reason. [An engaging recount of all this has been given by
Thorne~\cite{27}]

Due to general relativistic effects, the adiabatic exponent $\Gamma$ in the equation of
state is modified (increased) such that  $\Gamma > \frac{4}{3}$, is now the 
condition on $\Gamma$ for
the structure to be stable otherwise to collapse. Of course, such a collapse is what
must happen for the mammoth structure set out in Sec.~\ref{sect2}; a result that we certainly
want in our model of the large scale structure.

We can see how this happens from a heuristic calculation within the Newtonian
approximation which we have employed. The general relativistic effects on a massive
structure are to modify (strengthen) the gravitational attraction. We can model
that with a Newtonian potential modified as $1/R^{(1+\delta)}$.
 Now, balancing the pressure
$P \propto \rho^\Gamma$ against that due to the enhanced gravitational attraction,
$P_G \propto G^\prime \rho M/R^{(1+\delta)}$, [$G^\prime$ is a suitably adapted Newton's
constant], we have
%%%%%%%%%%%%%%%%%%%%%%%%%%%%%%%%%%%%%%%%%%%%%%%%%%%%%%%%%%%%%%%%
\begin{equation}
M \propto \rho^x \hspace*{1.0cm} {\rm where}\ \ x = 
\frac{(3\Gamma - 4 - \delta)}{(2-\delta)} .
\end{equation}
%%%%%%%%%%%%%%%%%%%%%%%%%%%%%%%%%%%%%%%%%%%%%%%%%%%%%%%%%%%%%%
The condition for equilibrium of a stellar structure is $\frac{dM}{d\rho} \ge 0$, 
which requires $\Gamma  \ge (4+\delta)/3$. So the mammoth, ultra-radiation dominated structure
with $\Gamma= 4/3$ as in Sec.~\ref{sect2} is unstable under enhanced gravitational 
attraction and must collapse.
Furthermore, we recognised from the outset, two other indicators of general relativistic 
effects. With the $M_J$ and $R_J$ for the structure in Sec.~\ref{sect2}, the general
relativity parameter $\epsilon$ is,
%%%%%%%%%%%%%%%%%%%%%%%%%%%%%%%%%%%%%%%%%%%%%%%%%%%%%%%%
\begin{equation}
\epsilon  =  \frac{G M_J}{R_J\ c^2} \simeq 1 ,
\end{equation}
%%%%%%%%%%%%%%%%%%%%%%%%%%%%%%%%%%%%%%%%%%%%%%%%%%
and the Hubble radius at this era, the dawn of structure ($z_r =1100$), is comparable
to the Jeans radius $R_J$ of the structure. For this we used the age of the current universe,
$t_u = 13.8\times 10^9$ yrs, and thus the age of the universe at this red-shift is
%%%%%%%%%%%%%%%%%%%%%%%%%%%%%%%%%%%%%%%%%%%%%%%%%%%%%%
\begin{equation}
t_r = t_u/z_r^\frac{3}{2} = 3.78.....\times 10^5 {\rm yrs}.
\end{equation}
%%%%%%%%%%%%%%%%%%%%%%%%%%%%%%%%%%%%%%%%%%%%%%%%%%%%%%%%%%%%%

Much interesting general relativistic work can be done on the model we propose
in Sec.~\ref{sect2} [some which we have undertaken but is as yet unpublished]. However, 
the essential
results and features of the model obtained from our Newtonian calculations allow
us to present a clear picture of the model's principle properties.

We believe these main results and features would endure under general relativistic 
scrutiny. Such is not unusual and we can think, straight away, of two such
cases. Oppenheimer and Volkoff solved the general relativistic hydrostatic equilibrium 
equation for a neutron star and found its mass to be approximately 1 $M_\odot$,
which is what one obtains from the Newtonian calculation. Only the pre-factor (1) is
slightly bigger, but the same scaled $M \propto (1/\alpha_G)^\frac{3}{2} H$ is found.
In our study~\cite{28,29} of the Jeans
stability condition in an expanding Robertson-Walker universe, we found that the
essential features of the non-relativistic Newtonian theory were preserved, e.g. the
same Jeans condition, albeit slightly modified due to the expansion, that we have
employed in Sec.~\ref{sect2}.

\section{The Model and The Scaling Law}
\label{sect5}

We envision that, at the era we call the dawn of structure, just prior
to $z_r = 1100$, the mammoth structure we presented and studied, in
Sec.~\ref{sect2} developed. Upon its collapse, it formed into an equally 
mammoth, ultra-planar structure, composed of ultra-degenerate neutrons as 
studied in Sec.~\ref{sect3}.
As we showed, this structure has just the Mass which scales in just the
right (same) way as the $M_J \propto \left(\frac{1}{\alpha_G}\right)^2 H$. 
And it has just the right (same) radius as the $R_J \propto
\left(\frac{1}{\alpha_G}\right) \lambda_H$.
Furthermore, importantly, it therefore has the right scaling law,
                  $M_J = \sigma_J R_J^2$.
This is just the right starting point to develop the ultimate scaling
law for the large scale structure of the universe.
 
We note that the collapse time for a structure can be estimated
 by its free-fall time, $t_{\rm ff} \propto (G \rho)^{-\frac{1}{2}}$. This time
 scales like $R^\frac{3}{2}(z)$, just as the age of the universe, 
$t_u = H^{-1}(z)$, [H(z) here is the Hubble parameter at red-shift, $z$], does at any 
 era of the expansion of the universe. Thus, we envision that any 
 such ultra-radiation dominated structures that formed earlier than
 the $z_r = 1100$ era would have already collapsed and disappeared, 
 with their constituent particles and photons passed back into the 
 universe. 

We continue to envision, that as this planar structure fragments, as it
must, as the universe expands into the era following $z_r = 1100$, 
the protons and electrons are liberated by the $\beta$-decay of the neutrons, 
the process of atomic formation takes place, followed by the process of stellar
formation, and then by the ultimate formation of the assemblies that
are the large scale structure.
 
Now, we propose that as these assemblies are formed with the expansion
of the universe, the planar nature of the structure that we have shown in
Secs.~\ref{sect2} and \ref{sect3} which we see as the progenerators of the 
large scale structure of the universe, persists, that is it is imprinted on 
the large scale structure, and that they will satisfy the same form of 
Mass-Radius scaling law, but now
%%%%%%%%%%%%%%%%%%%%%%%%%%%%%%%%%%%%%%%%%%%%%%%%%%
\begin{equation} 
M = \sigma R^2
\end{equation}
%%%%%%%%%%%%%%%%%%%%%%%%%%%%%%%%%%%%%%%%%%%%%%%%%%%%%%%%%%%%%%%%%%%%% 
where                    
%%%%%%%%%%%%%%%%%%%%%%%%%%%%%%%%%%%%%%%%%%%%%%%%%%%%
\begin{equation} 
\sigma = \frac{\sigma_J}{(z_r)^2} = \frac{M_J}{(R_J z_r)^2} = 0.0142...\ \ {\rm g/cm}^2
\label{sigeq}
\end{equation}
%%%%%%%%%%%%%%%%%%%%%%%%%%%%%%%%%%%%%%%%%%%%%%%%%%%%%%%%%%%%%% 
where $\sigma$ has scaled with the expansion of the universe
and  where $M$ and $R$ are the mass and radius of any of the assemblies that
constitute the large scale structure.

The two prime motivating insights presented in Sec.~\ref{sect1}, which are
supported by all that emerged in the model studies in the previous sections, 
are embodied in this ultimate result for the proposed Mass-Radius scaling 
law of the large scale structure of the universe, which has been our 
indefatigable quest.
 
We can cast this scaling law in its final, elegant, form, as,
%%%%%%%%%%%%%%%%%%%%%%%%%%%%%%%%%%%%%%%%%%%%%%%%%%%%%%%%%%
\begin{equation}
{\cal M} = {\cal R}^2
\end{equation}
%%%%%%%%%%%%%%%%%%%%%%%%%%%%%%%%%%%%%%%%%%%%%%%%%%%%%%%%%%%
where ${\cal M} = M/M_o$, 
$M$ is the mass of an assembly,
$M_o = M_J = 6.42 ...\times 10^{18} M_\odot$,  and where
${\cal R} = R/R_o$, with $R_o = R_J(z_r) = 308$ Mpc.
We note that there are no adjustable parameters in this
scaling law.
These $M_o$ and $R_o$ are just right to scale the current observable large
scale structure. To see just how right all this is, we have chosen 
values of $R$
from 0.01 Mpc through to 308 Mpc and computed the $M$ values that this
universal scaling law gives. All values are listed in Table~\ref{tab2}. It
is seen that all the assemblies in the large scale structure are accurately
described.
%%%%%%%%%%%%%%%%%%%%%%%%%%%%%%%%%%%%%%%%%%%%%%%%%%%%%%%
\begin{table*}[th]
\caption{\label{tab2}
Large Scale Structure Scaling$^\star$:\ \   ${\cal M} = {\cal R}^2$ }
\begin{ruledtabular}
\begin{tabular}{ccc} 
 $R$ (Mpc) & \hspace*{1.6cm} $M/M_\odot$ \hspace*{1.6cm} &
\hspace*{1.6cm} Structure \hspace*{1.6cm} \\ 
\hline
0.01\  $-$\ 0.1 & 6.8 $\times 10^{9}$\ $-$\ 6.8 $\times 10^{11}$& Galaxy\\
\hline
0.5\ $-$\ 1 & 1.7 $\times 10^{13}$\ $-$\ 6.8 $\times 10^{13}$ & Group of Galaxies\\
\hline
1\ $-$\ 5 & 6.8 $\times 10^{13}$\ $-$\ 1.7 $\times 10^{15}$ & Cluster of Galaxies\\
\hline
10\ $-$\ 50 &6.8 $\times 10^{15}$\ $-$\ 1.7 $\times 10^{17}$ & Super-cluster of Galaxies\\
\hline
80 & 4.3 $\times 10^{17}$ & CfA2 Great Wall~\cite{30}\\
\hline
210 & 3.0 $\times 10^{18}$ & Sloan Great Wall~\cite{31}\\
\hline
308 &  6.4 $\times 10^{18}$ & Gigas
\end{tabular} 
\phantom{$^\star$}\\
\end{ruledtabular}
$^\star$
{\small These structures are very elongated with one dimension
their length being much larger than their thickness.\\
\indent $R$ is identified with half the length. They are planar-like
structures.}
%\end{ruledtabular}
\end{table*}
%%%%%%%%%%%%%%%%%%%%%%%%%%%%%%%%%%%%%%%%%%%%%%%%%%%%%%%%%%%%%%%
 
We leave it for a very energetic person to scour all the catalogues and
literature for the (${\cal M},{\cal R}$) values for a legion of the observed
astronomical objects in the large scale structure, and plot them carefully on a
$\log[{\cal M}]$ vs $\log[{\cal R}]$ graph. They should, within experimental
error, lie on a universal straight line with slope = 2!

The last structure in Table~\ref{tab2}, is the ultimate Great Wall that the
scaling law yields. We have christened this structure, Gigas, which is a word 
meaning `Giant'. It was originally used to describe the race of Gigantes
in Greek mythology. We'll discuss their possible existence along with
other predictions and suggestions in the final Section,
Sec.~\ref{sect6}.

\section{Predications and Suggestions}
\label{sect6}

The scaling law we propose predicts that there will a largest
structure - a Supreme Great Wall, the Gigas - with a mass
$M = M_o$ and a radius $R_o$  as given in  Sec.~\ref{sect5}. 
At the beginning of 2013, the Sloan Great Wall~\cite{31}
was the largest known  Great Wall and it was still well within the Gigas
predicted by the scaling law. Then Clowes {\it et al.}~\cite{32} found
a Huge Large Quasar Group [Huge-LQG] in the Sloan
Digital Sky Survey. It has a radius of  620 Mpc and a
mass of  6.1 $\times 10^{18} M_\odot$. 

Now there are two possibilities. One is that this H-LQG Wall
is not all one structure just as 
suggested~\cite{33} for  the Sloan Great Wall that it is a chance alignment of three 
smaller structures.
The other is that it is really one distinct structure being now the largest
we know. For the scaling law, the first possibility is readily
accommodated by retaining it just as is with the Gigas, the
supreme structure.
Interestingly, the second can be accommodate easily by an
ever so slight tweaking of the results in Sec.~\ref{sect5}. We find for
the H-LQG structure that its surface mass density is
$M/R^2 = .0033...$. This is still remarkably close to the value of 0.0142
for the proposed scaling law as seen in Sec.~\ref{sect5}. So a very
small tweaking of $\sigma$ from 0.0142 to 0.0033, or in the pure
scaled form by tweaking $M_o$ to 0.95 $M_o$ and $R_o$ to 2.01 $R_o$,
leaves the law intact. The only other numerical
effect is to change the values of $M$ in Table~\ref{tab2} by the
small factor of 0.235. In either case [and we are naturally
partial to the first possibility], the proposed scaling law
stands.
 
We now look for any possible astrophysical or cosmological
indications of the model for the progeneration of the structure
that we have presented.
 
Most intriguingly, a new study~\cite{34}  of the Andromeda galaxy
has revealed that more than one-half of
its co-rotating dwarf galaxies all lie in the same vast thin
plane. And, as intriguing, this plane has a defined orientation
with a similar situation with respect to the co-rotating dwarf
satellite galaxies of another member of the Local Group; the
vast polar structure (VPOS) around the Milky Way~\cite{35} [see
also Ref.~\cite{36}]. These observations are in conflict with conventional 
astrophysical theory which would have all these dwarf galaxies
formed and distributed essentially randomly. Are these new
puzzling observations the tell-tale reflection of the kind of
planar-structure formation we envision?
 
A striking place to look for what has occurred at the $z_r = 1100$
era is to look to, and into, the CMB. Here also, most intriguingly
puzzling new results in the CMB have been revealed. In the latest
Planck mission, results recently announced~\cite{37} are: 
\begin{enumerate}
\item
The fluctuations should be random. Their distribution is random, but the 
amplitudes of the fluctuations are not. In the large map constructed by the 
Planck mission, the fluctuations are a tiny bit brighter on one side than
they should be, while being a tiny bit dimmer on the other side. 
This is heralded as the universe being `lopsided', as was
already seen in WMAP and is now confirmed by Planck. A simple
conventional model of the universe says this should not occur. 
\item 
The power angular fluctuations on a slightly large scale are seen to be
unexpectedly different - albeit a small difference - from those on
the smaller scales. This is not what current wisdom would
anticipate. 

These distribution results for the fluctuations at large angles 
first observed by WMAP have excited much interest, even 
to the extent of some suggesting that new physics beyond 
the standard model, the LCDM, is required~\cite{38,39,40,41,42}.

We would really like to see what much larger scale
fluctuations reveal and anxiously await the future where such
observations may be possible. 
\item
Planck has confirmed WMAP
observation, in microwaves, of the CMB Big Cold Spot. This is a
mammoth spot in the CMB  which is colder than the background
The diameter of this spot is huge. To date there is no explanation
of this wondrous object. 
\end{enumerate}
Could any of these results be the
tell-tale reflection of the kind of planar progeneration of
structure and evolution as envisioned herein?
 
\section{Conclusions}
 
There are some rather far-flung contemplations,
much more so than already proposed, but which follow on
from the considerations given. If the collapse of the proto-structure in 
Sec.~\ref{sect2} occurred,
then might there have been an accompanying cosmical amplitude
acoustical shock wave?  Recall that the speed of sound in such a
structure is $c/\sqrt{12}$ and it might be  associated with  such an  
ultra-powerful 
effect that the acoustical
vibrations may still be evident. And even further-flung yet, could such
rattle the space-time fabric, rippling it so violently, as to have
imposed an additional acceleration [like what is called dark
energy] on the expansion of the universe?
 
Furthermore, as envisioned at z = 1100, such ripples in the  
spacetime fabric -metric fluctuations-  would have important 
ramifications for the polarization of the CMB radiation.  
It would imprint in a  very specific way as the radiation 
scattering off such a gravity wave effect is quite different from
that off of the density fluctuations.
This is particularly relevant with respect to the very recently reported evidence of
B-modes in the polarization pattern of the CMB~\cite{43}. 

Then, if following on, the ultra-degenerate, planar proto-structure  of 
neutron matter collapse as envisioned in
Secs.~\ref{sect3} and ~\ref{sect5}, might there not still be evident the
copious spate of neutrinos that would have been released
at this $z_r = 1100$ era?
 
    Finally we note the striking similitude.  Both WMAP~\cite{1} and PLANCK~\cite{37} 
observe that the mass density of the universe is the critical mass density,
%%%%%%%%%%%%%%%%%%%%%%%%%%%%%%%%%%%%%%%%%%%%%%%%%%%%%%5
\begin{equation}
      \rho_c = 3 H^2/(8 \pi G) ,
\end{equation}
%%%%%%%%%%%%%%%%%%%%%%%%%%%%%%%%%%%%%%%%%%%%%%%%%%%%%%%% 
where the present value~\cite{37} of Hubble's constant is, $H = 67$ km/sec/Mpc. 
Further. it is observed~\cite{37} that the combined percentage contribution to 
$\rho_c$ from  the baryon and dark matter components is 31.7\%.

    Therefore, the mass $M_m$ of the universe from matter is
%%%%%%%%%%%%%%%%%%%%%%%%%%%%%%%%%%%%%%%%%%%%%%%%%%%%%
\begin{equation} 
         M_m = 0.317 \times \frac{4\pi}{3} \times \rho_c \times \left(R_H\right)^3
\end{equation}
%%%%%%%%%%%%%%%%%%%%%%%%%%%%%%%%%%%%%%%%%%%%%%%%%%%%%%%%%%%%%%%% 
where the visible radius of the universe, the Hubble's radius, $R_H = c t_u$
with the age of the universe being $t_u = 13.8 x 10^9$ years.
    
Thus, with
%%%%%%%%%%%%%%%%%%%%%%%%%%%%%%%%%%%%%%%%%%%%%%%%%%%%%%%%%%%%%%%%
\begin{align}
  M_m &= 2.489 ... \times 10^{55} {\rm g}\\
  R_H &= 1.305 ... \times 10^{28} {\rm cm}
\end{align}
%%%%%%%%%%%%%%%%%%%%%%%%%%%%%%%%%%%%%%%%%%%%%%%%%%%%%%%%%%%%%%%%%%%%%%%%%
we find for the surface mass density, $\sigma_m$, of the universe
%%%%%%%%%%%%%%%%%%%%%%%%%%%%%%%%%%%%%%%%%%%%%%%
\begin{align} 
  \sigma_m &= M_m/\left(4\pi (R_H)^2\right)\\
           &= 0.0116...  {\rm g/cm}^2
\end{align}
%%%%%%%%%%%%%%%%%%%%%%%%%%%%%%%%%%%%%%%%%%%%%%%%%%%%%%%%%%%%
which compares strikingly to the value 0.0142 ..., Eq.~(\ref{sigeq}), we have for the 
scaling of each of the assemblies of the large scale structure of the universe.

\begin{acknowledgments}
We are grateful to our long time good friend and colleague, Ken Amos, for his invaluable 
support in carrying out the numerical calculations of the structure equations
in Sec.~\ref{sect3} and for his preparation of the paper for publication.
\end{acknowledgments}

%%%%%%%%%%%%%%%%%%%%%%%%%%%%%%%%%%%%%%%%%%%%%%%%%%%%%%%%%%

\appendix
\section{Appendix}
What kind of objects that might have formed just after the recombination era had
begun, just after what we called the dawn of structure in Secs.~\ref{sect2} and
\ref{sect3} was considered by  Ref.~\cite{44}. 
Here the radiation has decoupled from the particles which are now free to 
gravitationally 
condense on their own. It was found that such first forming objects were
globular clusters.

Their analysis relied upon the same Jeans stability condition that we employed
in Sec.~\ref{sect2}. It is, therefore, very easy to adapt our 
results to obtain those for this era and recover their results. What is more,
we find them anew in their appropriate
scaled forms.

In this era, due to the fact that the internal pressure of the condensing object is
due to the particles, the radiation playing no role having decoupled, we have
directly now for the speed of sound,
%%%%%%%%%%%%%%%%%%%%%%%%%%%%%%%%%%%%%%%%%%%%%%%%%%%%
\begin{equation}
v^2 = 9\ \left(\alpha_G\right)^\frac{1}{4} \left( \frac{c^2}{12}\right) .
\end{equation}
%%%%%%%%%%%%%%%%%%%%%%%%%%%%%%%%%%%%%%%%%%%%%%%%%%%%%%%%%%%%%
We find straight away,
%%%%%%%%%%%%%%%%%%%%%%%%%%%%%%%%%%%%%%%%%%%%%%%%%%%%%%%%%%%%%%
\begin{equation}
M =7.18..... \times 
\left( \frac{1}{\alpha_G}\right)^\frac{13}{8} H = 8.0.....\times 10^5\ M_\odot ,
\end{equation}
%%%%%%%%%%%%%%%%%%%%%%%%%%%%%%%%%%%%%%%%%%%%%%%%%%%%%%%%%%%%%
and 
%%%%%%%%%%%%%%%%%%%%%%%%%%%%%%%%%%%%%%%%%%%%%%%%%%%%%%%%%%%%%%
\begin{equation}
R = 0.73.....
\left(\frac{1}{\alpha_G} \right)^\frac{7}{8} \lambda_H =
4.31.....\times 10^{19}\ \ {\rm cm}.
\end{equation}
%%%%%%%%%%%%%%%%%%%%%%%%%%%%%%%%%%%%%%%%%%%%%%%%%%%%%%%%%%%%%
This is the globular cluster.

Note that, while the globular cluster is a 3-dimensional object, it also scales in
a most surprising way with its surface density, namely $M/R^2 = 0.86$, 
which while it is not the surface mass density of the scaling law
 for the large scale structure of the universe, it is remarkably close.
 Thus, while the globular cluster is a 3 dimensional structure, it
 nonetheless  has a Mass-Radius which scales very much like any
 of the assemblies in the large scale structure of the universe. This
 is very suggestive in the light of our model. Furthermore this is
 in stark contrast with a structure like a star which is a natural 3 dimensional
 object, but has its surface mass density many orders of magnitude
 different from the globular cluster. As evident in Table~\ref{tab1}, the sun
 has a surface mass density 
$\left(\frac{1}{\alpha_G} \right)^\frac{1}{2} H/a_o^2 \simeq 10^{12}$;
a massively different value.

%%%%%%%%%%%%%%%%%%%%%%%%%%%%%%%%%%%%%%%%%%%%%%%%%%%%%%%%%%

\bibliography{PRE-Frankel}

\end{document}